\documentclass[sigconf]{acmart}
\usepackage{booktabs} 
\usepackage[utf8]{inputenc}
\usepackage{adjustbox}
\usepackage{paralist}
\usepackage{graphicx}
\usepackage{subcaption}
\usepackage{multirow}
\usepackage{bbm}
\usepackage{amsmath}
\usepackage{amsfonts}
\usepackage{amssymb}
\usepackage{url}
\usepackage{color}
\usepackage{mathtools}
\usepackage{tkz-graph}
\usepackage{tikz}
\usetikzlibrary{backgrounds}
\usetikzlibrary{trees,positioning,arrows}
\usepackage{wrapfig}
\usepackage[ruled,linesnumbered,lined,boxed]{algorithm2e}
\usepackage{array}
\usepackage{bm}
\usepackage{acronym}
\usepackage{enumerate}
\usepackage[shortlabels]{enumitem}
\usepackage[colorinlistoftodos]{todonotes}

\AtBeginDocument{%
  \providecommand\BibTeX{{%
    \normalfont B\kern-0.5em{\scshape i\kern-0.25em b}\kern-0.8em\TeX}}}

\acrodef{GBS}{Generalized Binary Search}

\newcommand{\RQRef}[1]{\textbf{\hyperref[section:setup:rq#1]{RQ#1}}}

\setcopyright{acmlicensed}

\acmDOI{10.475/123_4}
\acmISBN{123-4567-24-567/08/06}
\acmYear{2020}
\copyrightyear{2020}
\acmPrice{15.00}

\begin{document}
\fancyhead{}

\title{An Empirical Study of Clarifying Question-Based Systems}

\author{Jie Zou}
\affiliation{%
 \institution{University of Amsterdam}
 \city{Amsterdam}
  \country{The Netherlands}}
\email{j.zou@uva.nl}

  \author{Evangelos Kanoulas}
\affiliation{%
  \institution{University of Amsterdam}
  \city{Amsterdam}
  \country{The Netherlands}}
\email{e.kanoulas@uva.nl}

  \author{Yiqun Liu}
\affiliation{%
  \institution{BNRist, DCST, Tsinghua University}
  \city{Beijing}
  \country{China}}
\email{yiqunliu@tsinghua.edu.cn}

\renewcommand{\shortauthors}{J. Zou et al.}

\begin{abstract}
Search and recommender systems that take the initiative to ask clarifying questions to better understand users' information needs are receiving increasing attention from the research community. However, to the best of our knowledge, there is no empirical study to quantify whether and to what extent users are willing or able to answer these questions.
In this work, we conduct an online experiment by deploying an experimental system, which interacts with users by asking clarifying questions against a product repository. We collect both implicit interaction behavior data and explicit feedback from users showing that: (a) users are willing to answer a good number of clarifying questions (11-21 on average), but not many more than that; (b) most users answer questions until they reach the target product, but also a fraction of them stops due to fatigue or due to receiving irrelevant questions; 
(c) part of the users' answers (12-17\%) are actually opposite to the description of the target product; while 
(d) most of the users (66-84\%) find the question-based system helpful towards completing their tasks. Some of the findings of the study contradict current assumptions on simulated evaluations in the field, while they point towards improvements in the evaluation framework and can inspire future interactive search/recommender system designs.
\end{abstract}




\maketitle

\section{Introduction}

One of the key components of conversational search and recommender systems ~\cite{christakopoulou2016towards,zou2019learning,zhang2018towards} is the construction and selection of good clarifying questions to gather item information from users in a searchable repository. Most current studies either collect and learn from human-to-human conversations ~\cite{chen2019towards, sun2018conversational, li2018towards}, or create a pool of questions on the basis of some "anchor" text (e.g. item aspects~\cite{zhang2018towards, bi2019conversational}, entities~\cite{zou2019learning, zou2020towardsa, zou2020towardsb, zou2018technology}, grounding text~\cite{qi2020stay, nakanishi2019towards}) that characterizes the searchable items themselves. Although the aforementioned works have demonstrated success in helping systems better understand users, most of them evaluate algorithms by the means of simulations which assume users are willing to provide answers to as many questions as the system generates, and that users can always answer the questions correctly, i.e. they always know what the target item should look like in its finest details. On the basis of such assumptions, their evaluations (e.g. ~\citet{zou2019learning, zhang2018towards, bi2019conversational}) focus on whether the system can place the target item at a high ranking position. To the best of our knowledge, there is no empirical study to quantify whether and to what extent users can respond to these questions, and the usefulness perceived by users while interacting with the system. 

In this paper we conduct a user study by deploying an online question-based system to answer the following research questions:
\begin{enumerate}[itemsep=1mm]
\item To what extent are users willing to engage with a question-based system?
\item To what extent can users provide correct answers to the generated questions?
\item How useful do users perceive while interacting with a question-based system?
\end{enumerate}

The study is repeated under two conditions: (a) the question-based system uses an oracle to always obtain the right answer to the questions asked, and (b) the system uses the user's answers, even if they are imperfect, in ranking items and choosing the next question to ask. 
We believe that answering these research questions can help the community design better evaluation frameworks and more robust question-based systems.

\begin{figure*}
  \includegraphics[width=1.5\columnwidth]{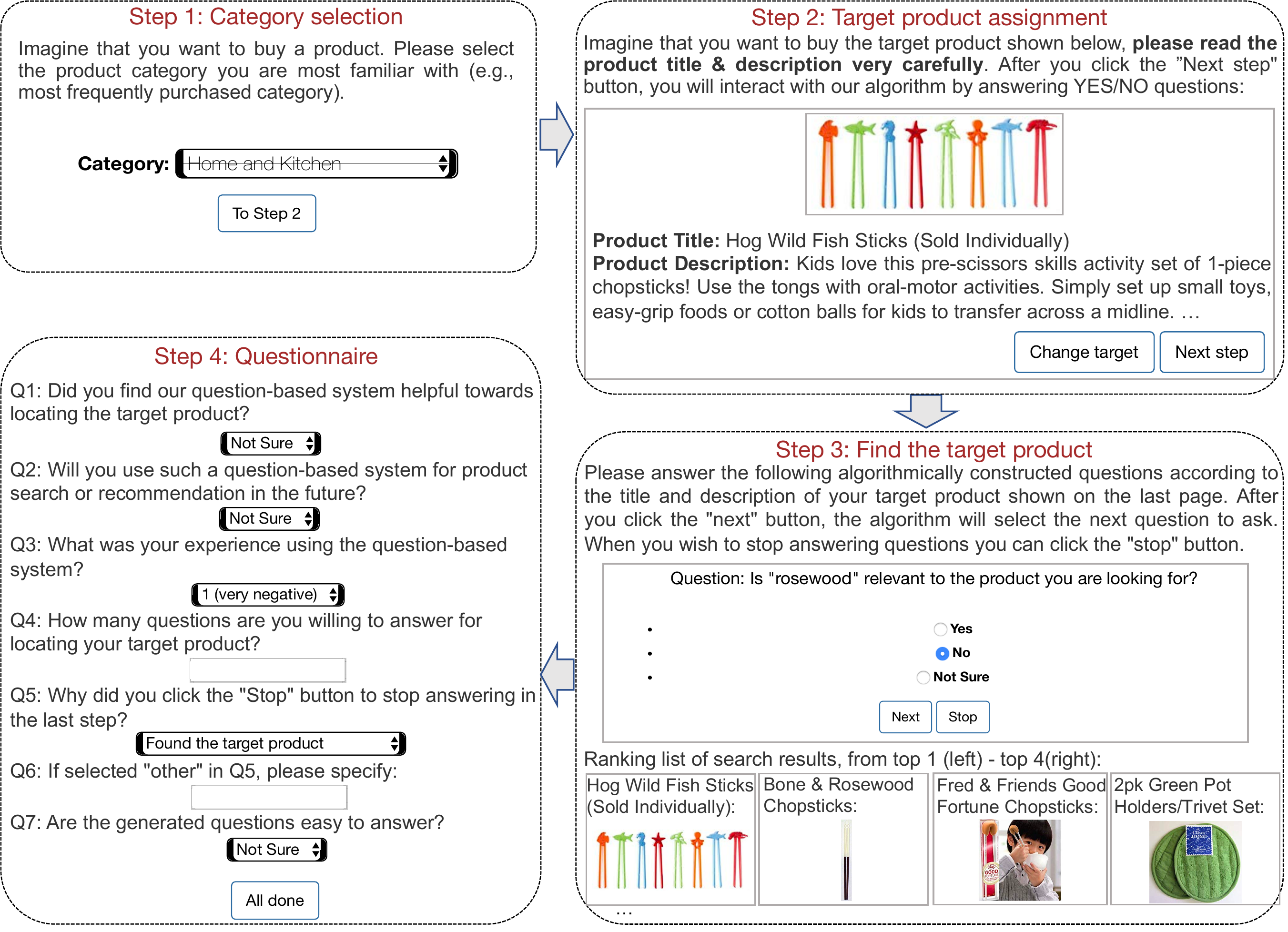}
  \captionsetup{font={small}}
  \caption{System architecture and main UI pages} 
  \label{fig:framework}
\end{figure*}

\section{Study Design}
In our study, the users interact with a question-based system in the domain of online retail. The user is answering questions prompted by the system with a ``Yes'', a ``No'' or a ``Not Sure'', in order to find a target product to buy. The architecture of our system is shown in Figure ~\ref{fig:framework}, with the user going through 4 steps.

\paragraph{Step 1: Category selection.}
In this step, the users select an Amazon category \footnote{Categories and dataset we used: http://jmcauley.ucsd.edu/data/amazon/links.html} that they feel most familiar with to fit their interests, e.g. a category from which they have purchased products before. 

\paragraph{Step 2: Target product assignment.}
We randomly assign a target product to the user from the selected category. The user is requested to read the title and the description of the product carefully. A picture of the product is also provided. Asking the users to carefully read the description simulates a use case in which the user really knows what she is looking for, as opposed to an exploratory use case. If the user is not familiar with the target product, or the description is not clear to her, the user can request a new randomly selected product.

\paragraph{Step 3: Find the target product.}
After the user indicates that the conversation with the system can start, the target product disappears from the screen and the system selects a question to ask to the user. The user needs to provide an answer on the basis of the target product information she read in the previous step. 
To help the user better understand her task of answering questions, an example target product along with an example conversation is also shown to the user.
Once the user answers the question, a 4-by-4 grid of the pictures of the top sixteen ranked products is shown to the user, along with the next clarifying question. The user can stop answering questions at any time when she wants to stop during her interaction with the system. 

To select what clarifying question to ask, a state-of-the-art algorithm~\cite{zou2019learning} is deployed to first extract important entities from each product description (e.g. product aspects) and construct questions in the form of "Is \textit{[entity]} relevant to the product you are looking for?". Then, it selects to ask the information-theoretically optimal question, that is the question that best splits the probability mass of predicted user preferences over items closest to two halves, and updates this predicted preference on the basis of the user's answer~\cite{zou2019learning}. 
In this work, we compare the results under two conditions: (a) the system updates the predicted preference using the correct answer, i.e. the answer which agrees with the description of the product, independent of the user's answer, and (b) the system updates its belief by using the user's noisy answers. Under the first condition, we study the user behavior under a perfect system from an information theoretical point of view, leading to a best-case analysis and conclusions, while under the second condition we study the user behavior when the system is getting confused and becomes suboptimal due to the user's mistakes. 

\paragraph{Step 4: Questionnaire.}
In this step users are asked a number of questions about their experience with the system for further analysis.

\section{Experiments and Analysis}
\subsection{Research Questions}
Our research questions revolve around the user engagement and perceived value of the system:
\begin{enumerate}[itemsep=1mm]
\item {\bf RQ1} Are users willing to answer the clarifying questions, how many of them, when do they stop and why, and how fast do they provide the answers?
\item {\bf RQ2} To what extent can users provide correct answers given a target product, and what factors affect this?
\item {\bf RQ3} How useful do users find the clarifying questions, what is their overall experience, and how likely is it to use such a system in the future?
\end{enumerate}

\subsection{Participants}
Prior to the actual study, we ran a pilot study with a small number of users, in a controlled environment, and iterated over the experimental design, and the user interface until no issues or concerns were reported. Then we considered two conditions. Under the first one the system used an oracle to obtain the correct answers to the questions it asked to the system. 
For the actual study 53 participants located in the USA were recruited through Amazon Mechanical Turk and 1025 conversations were collected. The participants were of varying gender, age, career field, English skills and online shopping experience. In particular, gender: 34 male, 19 female; age: 2 in 18-23, 8 in 23-27, 14 in 27-35, 29 older than 35 years old; career field: 22 in science, computers \& technology, 8 in management, business \& finance, 7 in hospitality, tourism, \& the service industry, 3 in education and social services, 2 in arts and communications, 2 in trades and transportation, 9 did not specify their career field; English skills: all of them were native speakers; online shopping experience: 44  were mostly shopping online, 9  did online shopping once or twice per year. 
Under the second condition the system actually used the users answers, with 48 users participating in this one, also with varying demographic and skills characteristics. 1833 conversations are collected for these 48 users. 
Participants were paid 2.5 dollars to complete the study. Also, we only engaged Master Workers \footnote{High performing workers identified by Mechanical Turk who have demonstrated excellence across a wide range of tasks.}, filtered out those users who spent less than 3 seconds on reading the product title and descriptions, and users who gave random answers (\textasciitilde 50\% correct/wrong), for quality control. 
\begin{figure}
  \begin{subfigure}{0.25\textwidth}
    \includegraphics[width=\columnwidth]{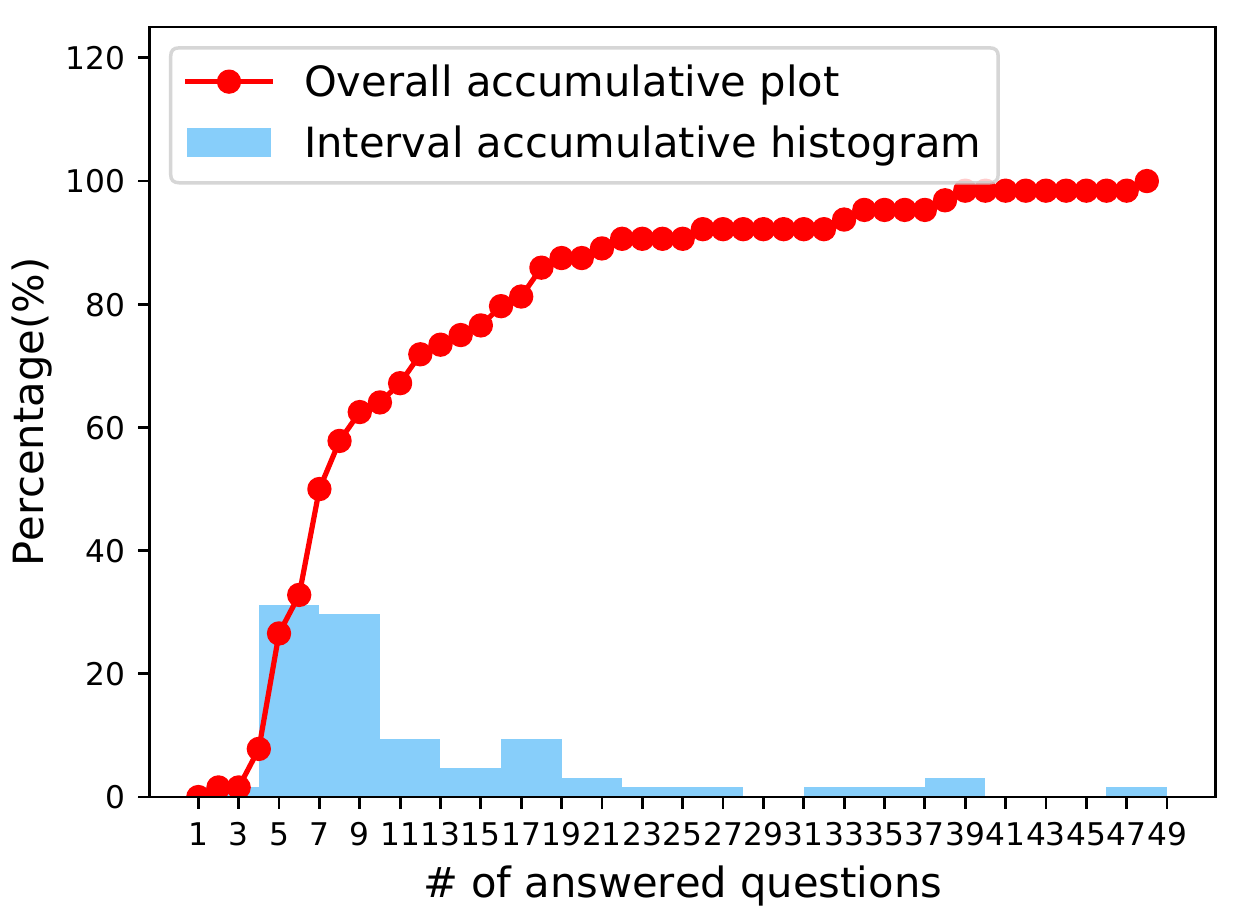}
  \end{subfigure}
  \begin{subfigure}{0.14\textwidth}
    \includegraphics[width=\columnwidth]{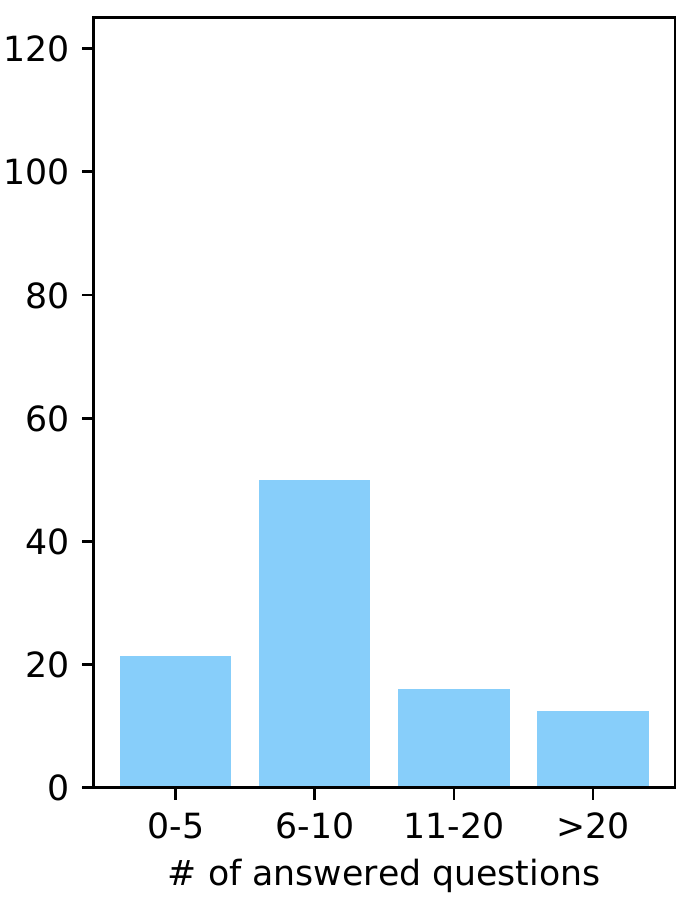}
  \end{subfigure}
  \captionsetup{font={small}}
  \caption{The number of questions the users actually answered in the system (left) and declared in the exit questionnaire (right). In the system, the average number of answered questions per product is 11.4, and 70.3\% of users answered 4-12 questions per product. In the questionnaire, 50\% users are willing to answer 6-10 questions.} 
  \label{fig:rq1}
\end{figure}
\subsection{~\RQRef{1} User Willingness to Answer Questions}
To answer ~\RQRef{1}, we attempt to answer the following sub-questions: 
\begin{inparaenum}
\item Are users willing to answer the system's questions?
\item How many questions the users are willing to answer?
\item When do they stop and why?
\item How fast are they able to answer?
\end{inparaenum}

In \RQRef{1}, we first investigate whether users are willing to answer the system's questions and how many of them, both by observing the actual number of questions users answered when interacting with the question-based system and what they declared at the exit questionnaire. The findings under the oracle condition are summarized in Figure ~\ref{fig:rq1}, with the left subfigure depicting the actual number of questions answered by the users when interacting with our question-based system, while the right subfigure depicting the number of questions the users declared they are willing to answer in the exit questionnaire. In the left subfigure of Figure ~\ref{fig:rq1}, the red line represents the accumulated percentage of users willing to answer a certain number of questions; the light blue histogram reflects the percentage per number of questions.
The results in Figure ~\ref{fig:rq1} show that users answer a minimum of 2 and a maximum of 48 questions. The average number of questions answered per target product is 11.4, the median number is 7, and 70.3\% of users answered 4-12 questions per product, while at the exit questionnaire about 50\% of the users declare that they are willing to answer 6-10 questions. 

We also compare the afore-described statistics with those under the condition that the system updates its beliefs and hence chooses the next question and ranks items according to the actual user answers, however imperfect they might be. In this latter case we observe that the average number of questions answered per target product is 21, and the median number is 14, which is almost double than that of using an oracle. We do not provide these plots due to space limitations. We hypothesize (and we later confirm that in the exit questionnaire) that this is because our users really try to locate the target product and they go as far as it takes to make that happen or get frustrated; however their noisy answers confuse the algorithm and it takes longer to bring the target product to the top of the recommendation list.

\begin{figure}
    \includegraphics[width=0.6\columnwidth]{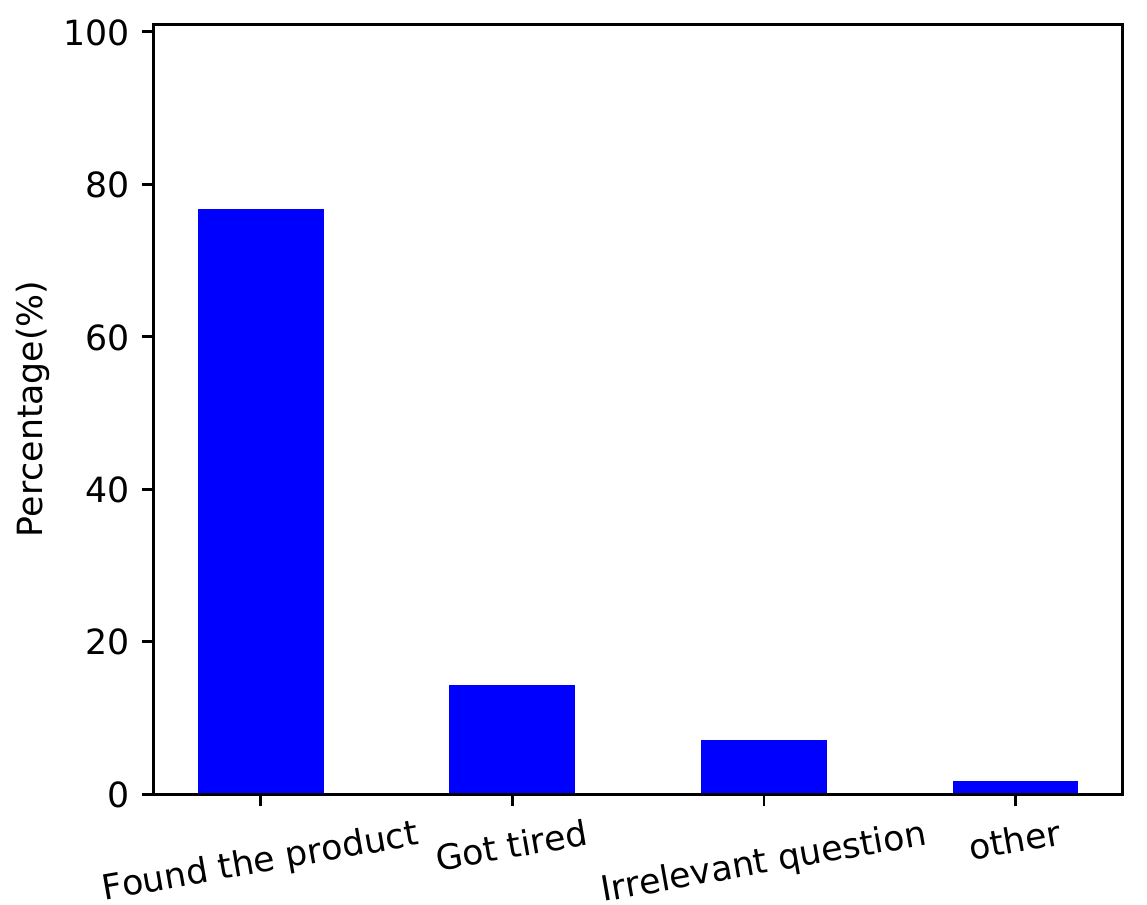}
    \captionsetup{font={small}}
  \caption{The reasons for stopping answering questions. Most of users stop answering questions after they find their target products.} 
  \label{fig:rq1_3}
\end{figure}

\begin{figure}
  \begin{subfigure}{0.08\textwidth}
    \includegraphics[width=\columnwidth]{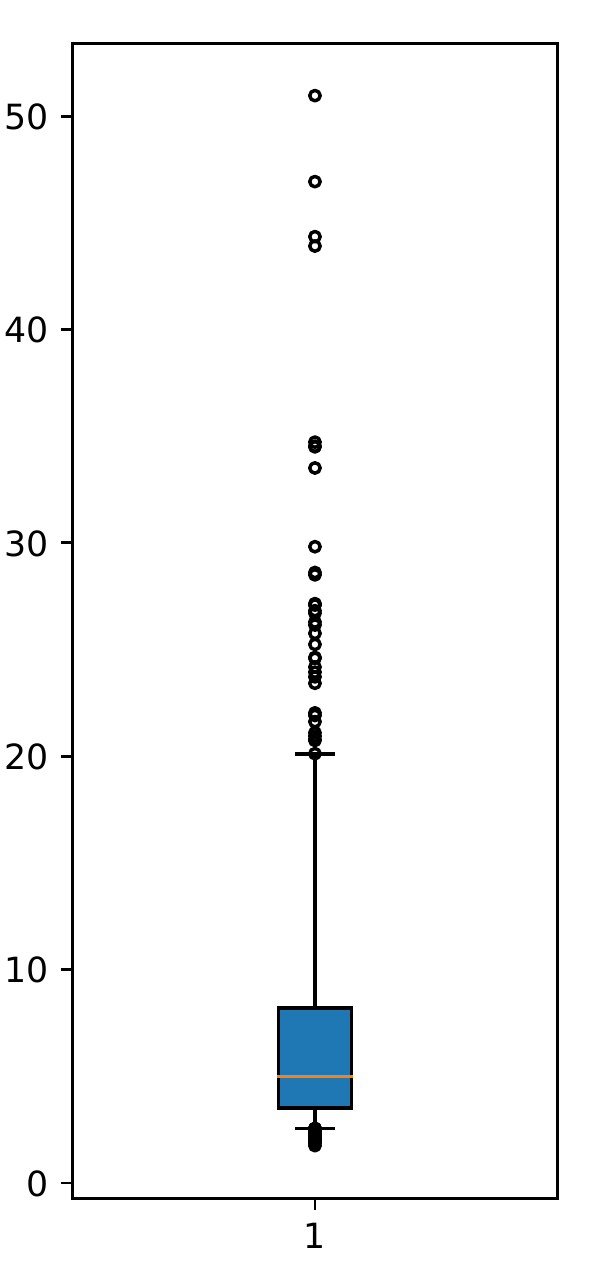}
    \captionsetup{font={small}}
    \caption{Box plot} \label{fig:rq1_4a}
  \end{subfigure}
    \begin{subfigure}{0.26\textwidth}
    \includegraphics[width=\columnwidth]{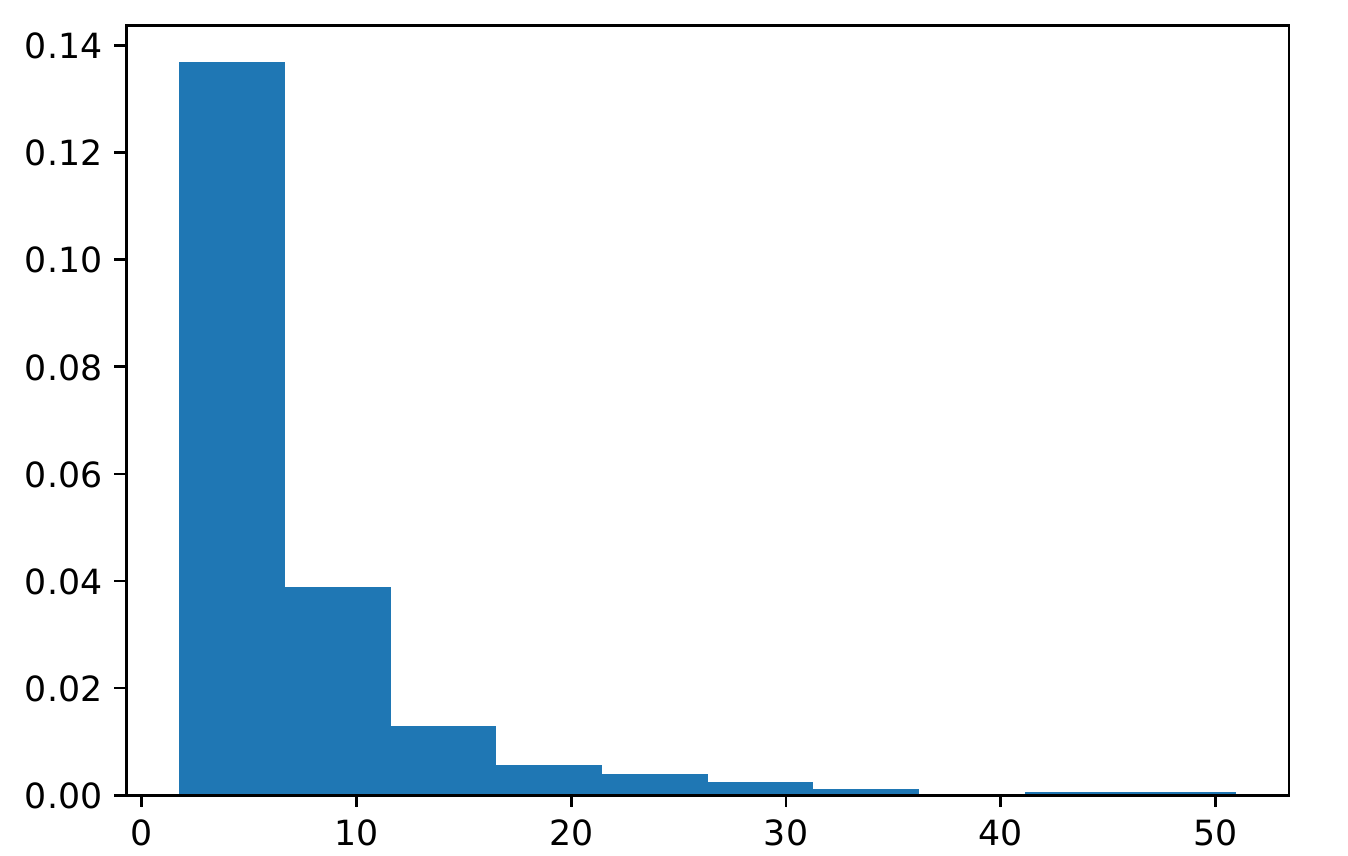}
    \captionsetup{font={small}}
    \caption{Histogram of time spent} \label{fig:rq1_4b}
  \end{subfigure}
  \captionsetup{font={small}}
  \caption{Time spent per question by a user in order to provide an answer. The average time for answering one question is 7.1 seconds.}
  \label{fig:rq1_4}
\end{figure}

Further, we explore why users stop answering questions. 
Users could select one out of six answers during the exit questionnaire: "The target product was found", "A similar product was found", "I got tired of answering questions", "I could not answer the questions", "The questions asked were irrelevant", and "Other". The results under the oracle condition in Figure ~\ref{fig:rq1_3} show that while a small percentage of users stop due to fatigue (14\%) or due to irrelevant questions being asked (7\%), the big majority of users (77\%) stop because they located the target product. Under the second condition of imperfect answers most users also stop answering questions because they found the target products  (38\%), but other reasons are more prominent such as fatigue (34\%), or receiving irrelevant questions (22\%).

We then analyze how quick are the users in answering questions. Figure ~\ref{fig:rq1_4a} shows a box-plot of the time spent per answer, while ~\ref{fig:rq1_4b} better demonstrated the distribution. From the results, we observe that the minimum time for answering one question is 1.75 seconds, the average time is 7.1 seconds, and the median time is 4.98 seconds. 86.5\% of the users spent from 1.75s to 11.59s. Despite a median time of 5 seconds to answer a question, in the exit questionnaire 98\% of the users indicate that the system's questions are easy to answer.
When using the user's imperfect answers the average time for answering a question is 6.22 seconds, while the median time is 4.1 seconds, which is similar to that using an oracle.

\begin{figure*}
  \begin{subfigure}{0.3\textwidth}
    \includegraphics[width=\columnwidth]{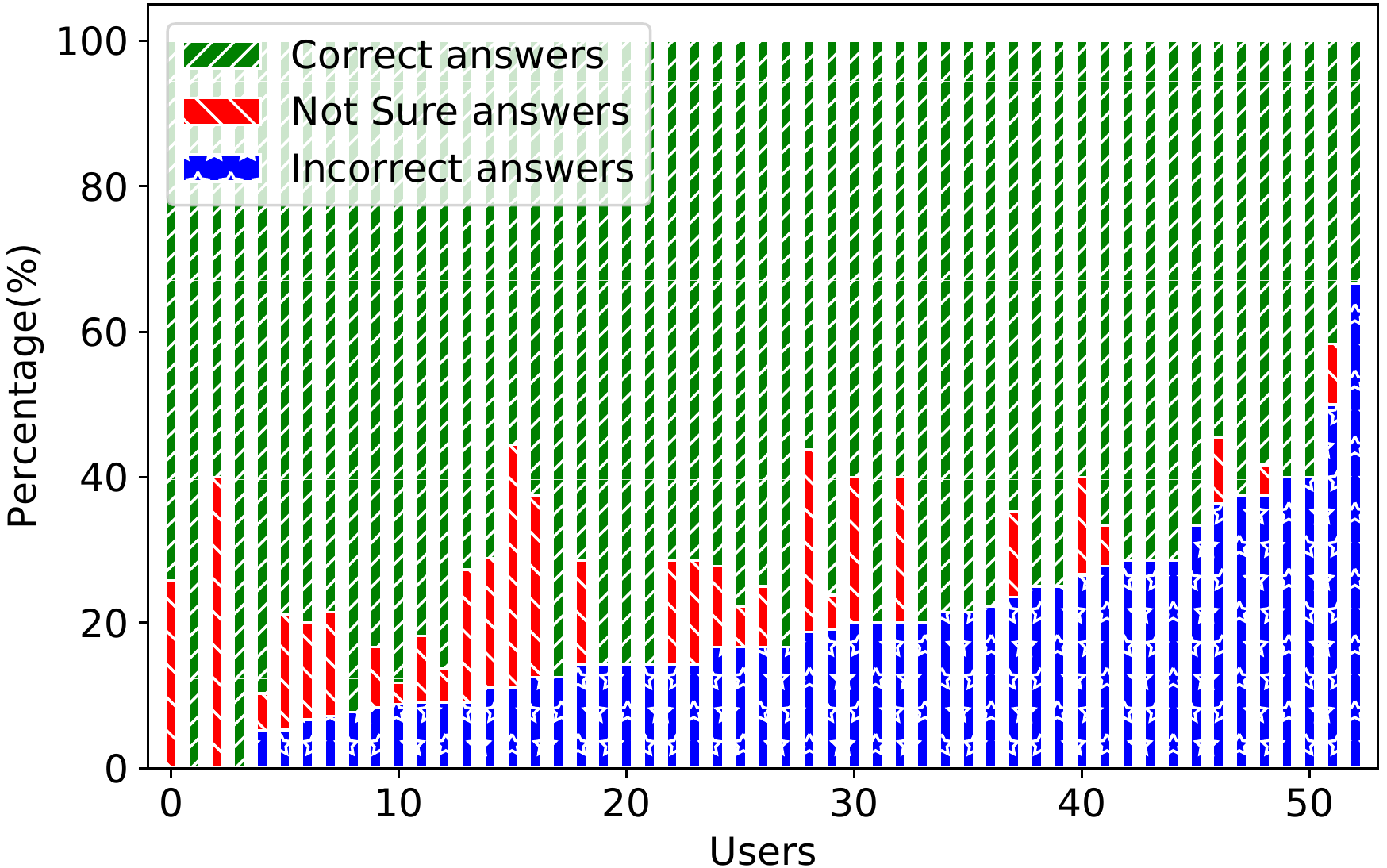}
    \captionsetup{font={small}}
    \caption{Results under different users}   \label{fig:rq2_2}
  \end{subfigure}
  \begin{subfigure}{0.3\textwidth}
    \includegraphics[width=\columnwidth]{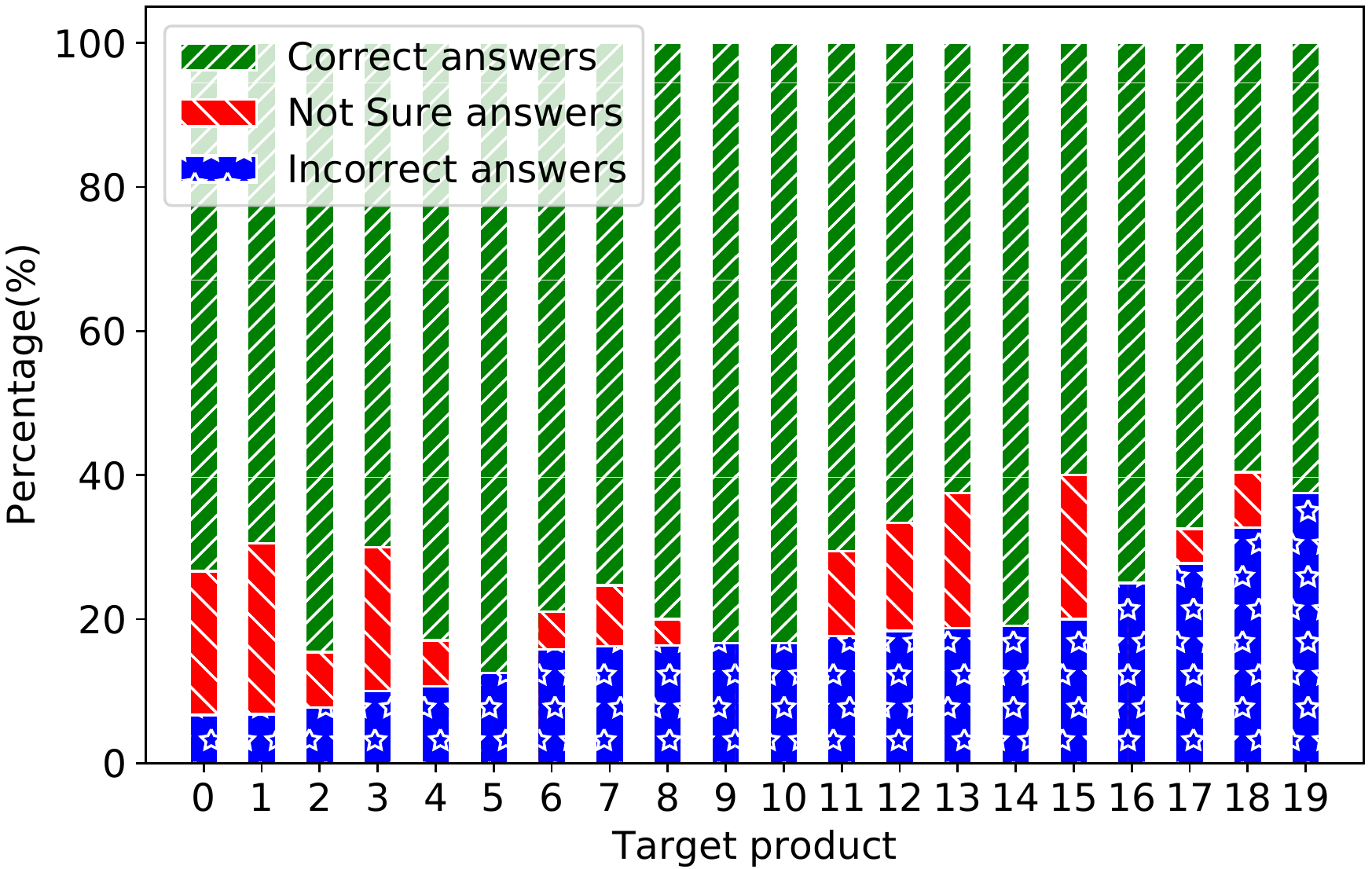}
    \captionsetup{font={small}}
     \caption{Results under different target products}   \label{fig:rq2_3}
  \end{subfigure}
      \begin{subfigure}{0.3\textwidth}
    \includegraphics[width=\columnwidth]{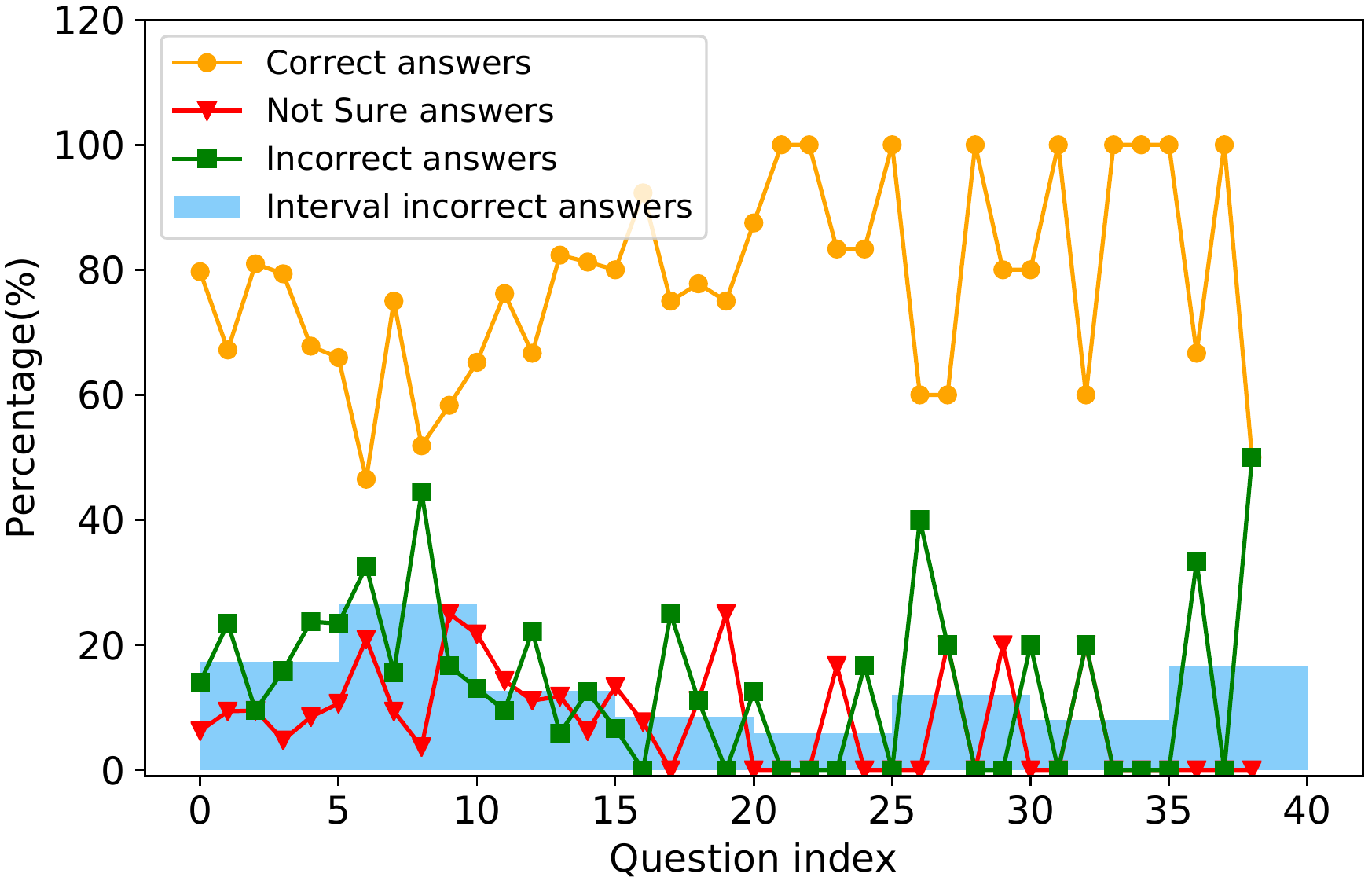}
    \captionsetup{font={small}}
    \caption{Results under different question index}   \label{fig:rq2_4}
  \end{subfigure}
    \begin{subfigure}{0.6\textwidth}
    \includegraphics[width=\columnwidth]{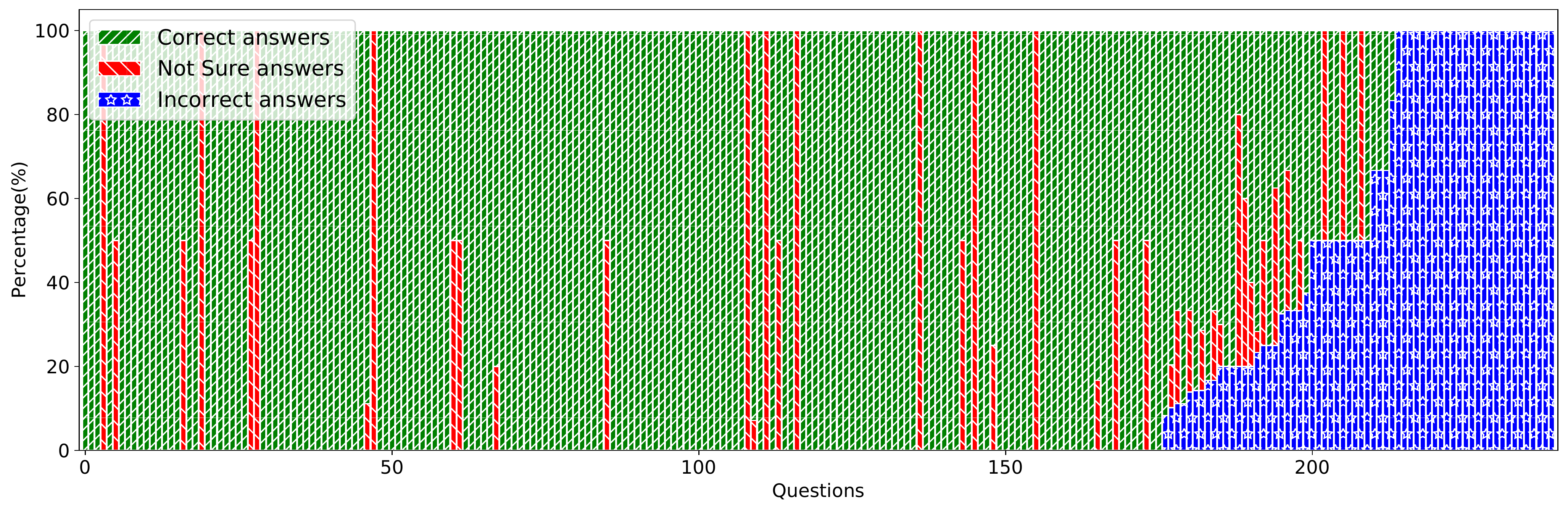}
    \captionsetup{font={small}}
    \caption{Results under different questions}   \label{fig:rq2_6}
  \end{subfigure}
    \begin{subfigure}{0.3\textwidth}
    \includegraphics[width=\columnwidth]{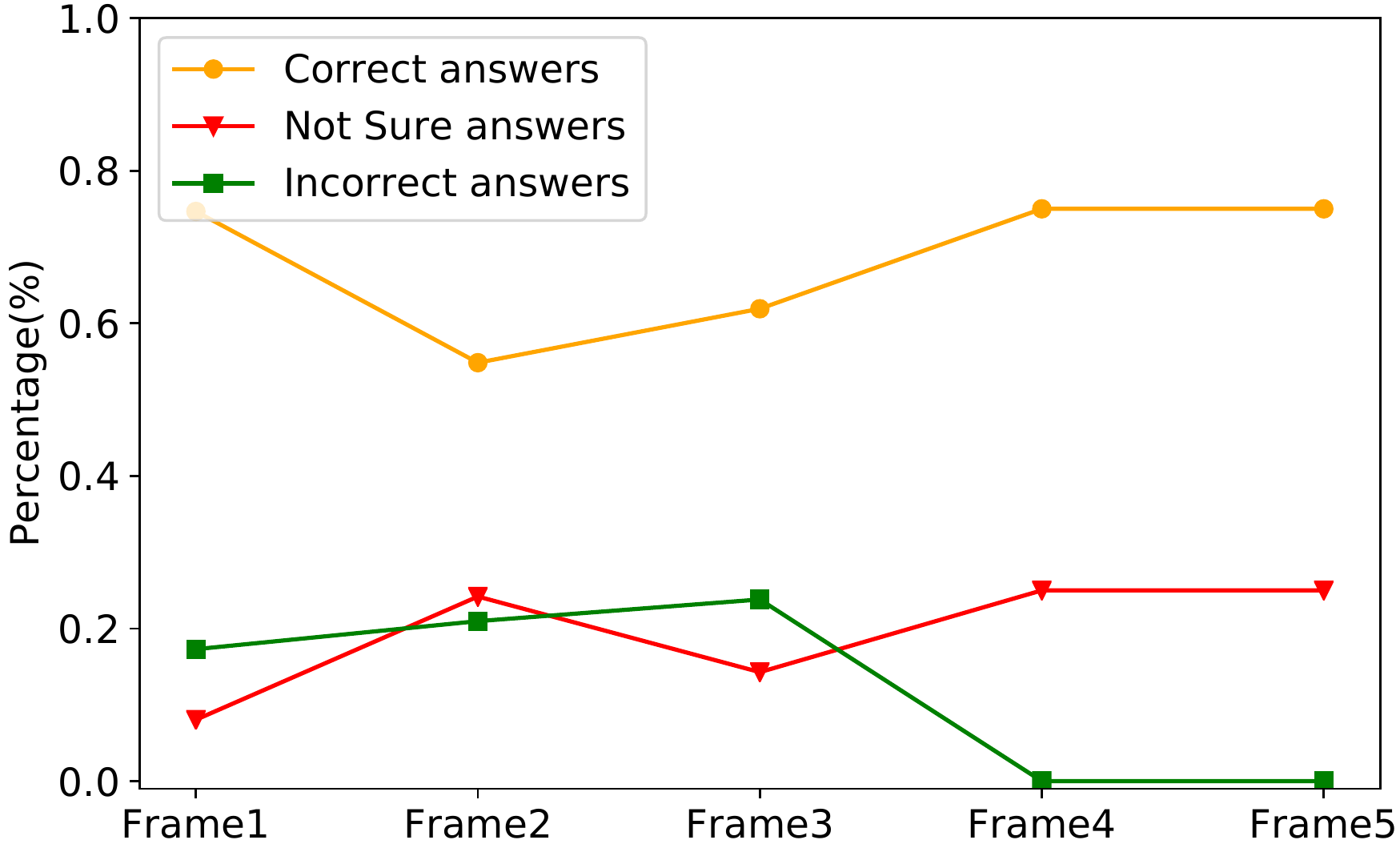}
    \captionsetup{font={small}}
    \caption{Results under different time windows} \label{fig:rq2_5}
  \end{subfigure}
  \captionsetup{font={small}}
  \caption{The percentage of correct answers, ``not sure'' answers, which cannot be classified, and incorrect answers. (a) The \% varies per user; (b) The \% varies across different target products; (c) The \% remains stable through out the conversation; (d) The \% varies per question, with only few questions receiving most of the incorrect answers; (e) the \% of incorrect answers decreases with more time spent.}
  \label{fig:RQ2}
\end{figure*}

\subsection{~\RQRef{2} User Answers Noise}

In \RQRef{2}, we first explore to what extent can users provide correct answers. As one can observe in Table ~\ref{table:rq2_0}, users provide correct answers 73.1\% of the time, they are not sure 9.6\% of the time and they are wrong 17.3\% of the time. Under the imperfect user answer setup, the afore-described percentages are 78.3\%, 9.5\%, and 12.2\%, respectively.

We then explore what features affect the percentage of incorrect answers. We do that under both setups but we only report the oracle setup given that numbers are very close across the two setups.

In particular, we first investigate whether the percentage of incorrect answers is different for different users. The results in Figure ~\ref{fig:rq2_2} show the percentages of correct, ``not sure'', and incorrect answers vary across users. It might be because of the varied knowledge of different users. 
There is a couple of users who provide a high percentage of incorrect answers, this might be because of the crowdsourcing nature of the experiment.

Further, we explore whether the percentage of correct answers differs across target products. The results are shown in Figure ~\ref{fig:rq2_3}. By observing Figure ~\ref{fig:rq2_3}, we conclude that the percentages of incorrect answers also vary across target products, but not as much as they vary across users.

The percentages of correct, ``not sure'', and incorrect answers for different questions asked by the system are shown in Figure ~\ref{fig:rq2_6}. Here we observe some dramatic differences across questions, with a smaller subset of questions receiving almost always incorrect answers. This might be because some questions are more ambiguous than others. This finding suggests improvements of question-based systems in multiple directions. For instance, one can try to improve the question pool by considering different question characteristics, or one could develop question selection strategies that also account for the chance of user providing the wrong answer. 

Further, we explore whether the percentage of incorrect answers is correlated to the question index, or whether it remains stable throughout the conversation. The results are shown in Figure ~\ref{fig:rq2_4}, where the lines show the average percentages of correct, ``not sure'', and incorrect answers as a function of the question index within the conversation, while the histogram shows the average incorrect answer percentages of a sliding window. From Figure ~\ref{fig:rq2_4}, it can be observed that the percentages of correct answers, ``not sure'' answers, and incorrect answers fluctuate, but in principle they remain at similar levels throughout the conversation. 

Last, we explore whether the percentage of incorrect answers is correlated to the time spent to give the answers. The results within different time intervals are shown in Figure ~\ref{fig:rq2_5}. We divide the time spent per question (1.75s - 50.96s) into 5 equal non-overlapping buckets (or Frames). Specifically, Frame 1, Frame 2, Frame 3, Frame 4, and Frame 5 are 1.75-11.59s, 11.59-21.43s, 21.43-31.28s, 31.28-41.12s, and 41.12-50.96s, respectively. From Figure ~\ref{fig:rq2_5}, we see the percentage of incorrect answers decreases with more time spent. Also, we calculate the time spent when users are giving a correct answer, a ``not sure'' answer, and an incorrect answer, with the averages being 6.59s, 10.81s, and 7.12s respectively, and the median 4.65s, 8.20s, 5.06s respectively. This suggests that users usually spent more time when they are not sure about the answers, but almost the same time when they are right or wrong about a question.

\begin{table}
\captionsetup{font={small}}
\caption{The \% of correct, ``not sure'', and incorrect answers.}
\label{table:rq2_0}
\centering
  \small
\begin{tabular}{c|c|c|c|c|c}
\toprule
 \multicolumn{6}{c}{Oracle}\\
\hline
Correct & 73.1\%& Not sure & 9.6\% &Incorrect & 17.3\%\\
\hline
\multicolumn{6}{c}{Imperfect user}\\
\hline
Correct & 78.3\%& Not sure & 9.5\% &Incorrect & 12.2\%\\
\bottomrule
\end{tabular}
\end{table}

\begin{figure}
  \begin{subfigure}{0.19\textwidth}
    \includegraphics[width=\columnwidth]{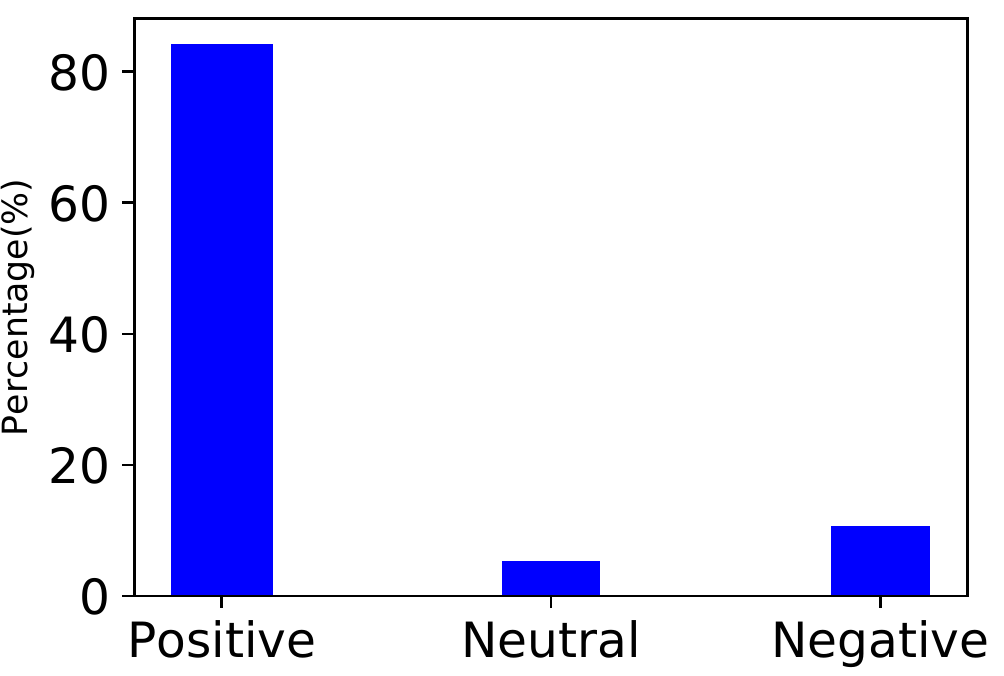}
    \captionsetup{font={small}}
    \caption{} \label{fig:rq3a}
  \end{subfigure}
  \begin{subfigure}{0.14\textwidth}
    \includegraphics[width=\columnwidth]{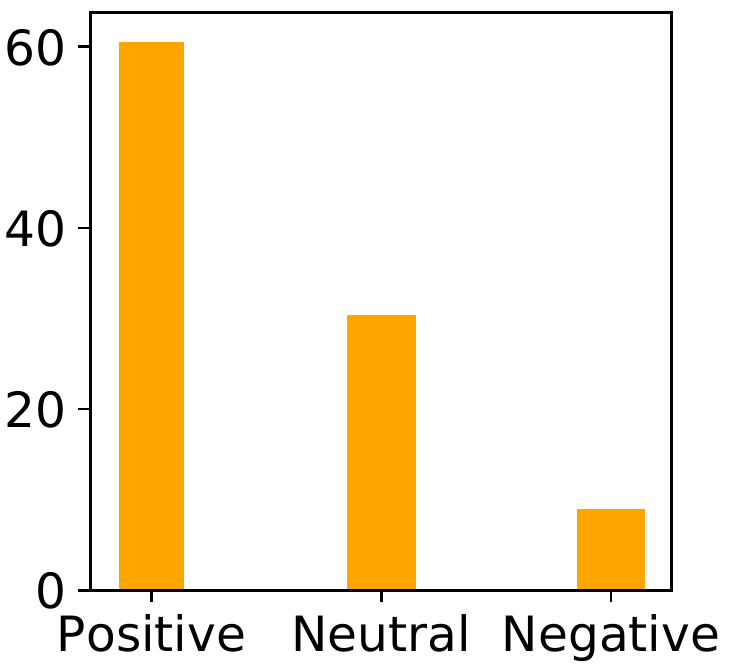}
    \captionsetup{font={small}}
     \caption{} \label{fig:rq3b}
  \end{subfigure}
    \begin{subfigure}{0.13\textwidth}
      \includegraphics[width=\columnwidth]{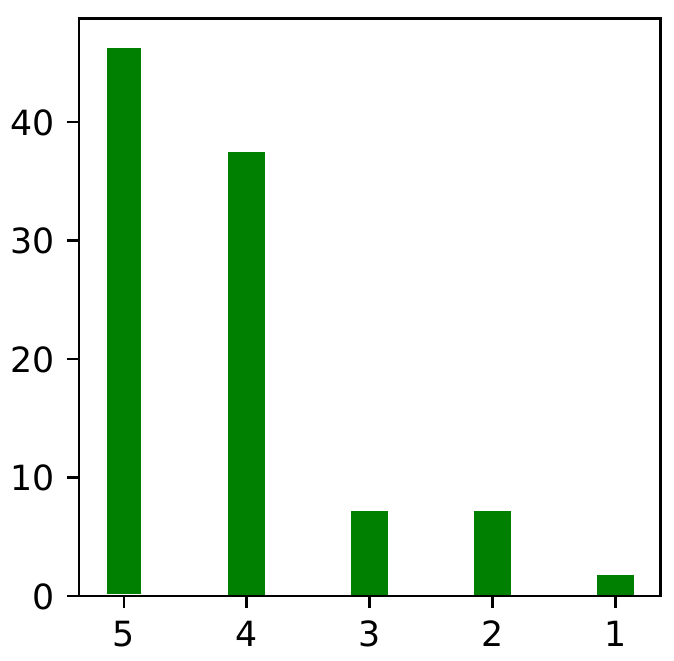}
      \captionsetup{font={small}}
       \caption{} \label{fig:rq3c}
  \end{subfigure}
  \captionsetup{font={small}}
  \caption{User perceived helpfulness. (a) Is the question-based system helpful; (b) will you use the question-based system in the future; (c) ratings. Most users are positive towards question-based systems.} 
  \label{fig:rq3}
\end{figure}

\subsection{~\RQRef{3} User Perceived Helpfulness}
Regarding ~\RQRef{3}, we explore how useful do users perceive while interacting with such a question-based system. We ask the user (a) whether they think the question-based system is helpful, (b) whether they will use such a system in the future, and (c) what their rating is for the system, ranging from 1 (very negative) to 5 (very positive). The results using oracle answers are shown in Figure ~\ref{fig:rq3}, in the three plots respectively.
From the results we collected, most users think the question-based system is helpful and they will use it in the future. Specifically, 83.9\% of users are positive about the helpfulness, 5.4\% are neutral, and 10.7\% are negative. Further, 60.7\% of users are positive about using such a system in the future, 30.4\% of users are neutral, and 8.9\% of users are negative. Regarding user ratings, the results show 46.5\% of 5-star ratings, 37.5\% of 4-star ratings, 7.1\% of 3-star ratings, 7.1\% of 2-star ratings, and 1.8\% of 1-star ratings. 84\% of the users gave a rating at least as high as a 4.

In the case of imperfect answers, 66\% of users are positive, 6\% are neutral, and 28\% are negative about being helped by the system. Regarding using such a system in the future 40\% of users are positive, 20\% are neutral, and 40\% are negative. Regarding ratings, 76\% of users gave the rating greater or equal to 3. Specifically, there is 22\% of 5's, 32\% of 4's, and 22\% of 3's, 16\% of 2's, and 8\% of 1's. Still, most users are positive towards conversational recommender. But we also observe that, when using user answers updating, users are less positive than that under oracle answers updating. It is therefore clear that the quality of the user answers affects the quality of the system questions and the overall user experience with the system.
 
\section{Conclusion and Discussion}
In this paper, we conduct an empirical study using a question-based product search system, to better understand users and gain insight into user behavior and interaction with such systems.
We deploy a state-of-the-art question-based system online and collect interactive log data and questionnaire data for analysis. We find that users are willing to answer a certain number of the system generated questions and stop answering questions when they find the target product, only if the questions are relevant and well-selected.

While users are able to answer these questions effectively, however we also observe that users provide incorrect answers at a rate of about 17\%, and this rate is affected mostly by some, still to be identified, question characteristics, while it also varies across users and products.
Last, most users are positive towards question-based systems, and think that these systems help them towards achieving their goals, although this feeling is weaker with systems not robust to imperfect answers. 
The take-home message, if there is one, is that current research should drop the assumption that users are happy to answer as many questions as the system generates and that all questions are answered correctly.

One limitation of this work is the isolated clarifying-based environment of the study. A more realistic experiment would require clarifying questions to be embedded in an existing environment, where the user is enabled to not only answer questions, but also reformulate her query or filter results by selecting pre-defined item attributes, and browse the results to the preferred depth. Also a mixed-initiative approach under which a system switches from asking questions, to understanding user searches, and combining the two is worth studying. Such a study is in our future plans.
A further limitation of this work is the fact that this was not an in-situ experiment but a simulation of a use case of a question-based system by involving crowd workers. Hence, the findings are as good as our simulation of a user looking for a target product. Furthermore, we cannot know whether by running the study for a long period of time the results would have been the same, or whether we are observing some novelty effects~\cite{kohavi2017online}.
Other factors, such as question quality, question format (e.g. yes/no or open questions), and noisy answers, may affect the results, and studying therefore these factors in an A/B testing experiment would be beneficial. We also leave that for future work.

\bibliographystyle{ACM-Reference-Format}
\bibliography{bibfile} 

\end{document}